\documentclass[amsmath,11pt]{article}

\usepackage{amsfonts,amsmath,amssymb,graphics,epsfig}

\usepackage{appendix}

\date{\empty}

\begin{document}

\author{Christos G. Tsagas\\ {\small Section of Astronomy, Astrophysics and Mechanics, Department of Physics}\\ {\small Aristotle University of Thessaloniki, Thessaloniki 54124, Greece}}

\title{\bf On the Magnetic Evolution in Friedmann Universes and the Question of Cosmic Magnetogenesis}

\maketitle

\begin{abstract}
We analyse the evolution of primordial magnetic fields in spatially flat Friedmann universes and reconsider the belief that, after inflation, these fields decay adiabatically on all scales. Without~abandoning classical electromagnetism or standard cosmology, we demonstrate that this is not necessarily the case for superhorizon-sized magnetic fields. The underlying reason for this is causality, which confines the post-inflationary process of electric-current formation, electric-field elimination and magnetic-flux freezing within the horizon. As a result, the adiabatic magnetic decay is not a~priori guaranteed on super-Hubble scales. Instead, after inflation, large-scale magnetic fields obey a~power-law solution, where one of the modes drops at a rate slower than the adiabatic. Whether this slowly decaying mode can dominate and dictate the post-inflationary magnetic evolution depends on the initial conditions. These are determined by the evolution of the field during inflation and by the nature of the transition from the de Sitter phase to the reheating era and then to the subsequent epochs of radiation and dust. We discuss two alternative and complementary scenarios to illustrate the role and the implications of the initial conditions for cosmic magnetogenesis. Our main claim is that magnetic fields can be superadiabatically amplified after inflation, as long as they remain outside the horizon. This means that inflation-produced fields can reach astrophysically relevant residual strengths without breaking away from standard physics. Moreover, using the same causality arguments, one can constrain (or in some cases assist) the non-conventional scenarios of primordial magnetogenesis that amplify their fields during inflation. Finally, we show that our results extend naturally to the marginally open and the marginally closed Friedmann universes.
\end{abstract}

\section{Introduction}\label{sI}
The origin of cosmic magnetism remains an essentially open question despite the efforts and the established widespread presence of magnetic ($B$) fields in the universe~\cite{K}-\cite{V}. Recent reports of the first ever detection of intergalactic fields, with strengths around $10^{-16}$~G, have added to the mystery~\cite{NV}-\cite{CBF}. Provided~they are verified, these claims also support the case for primordial magnetism~\cite{KKT,Wetal}. The~latter is an attractive proposition because it could potentially explain all the large-scale $B$-fields of the universe. Nevertheless, there are serious theoretical problems in producing such primordial fields. These mainly stem from the long standing belief that (conventional) magnetic fields in spatially flat Friedmann-Robertson-Walker (FRW) universes decay adiabatically throughout the evolution of these models and on all scales.

The structure of the magnetic fields in galaxies seems to support the galactic-dynamo idea~\cite{KA}-\cite{BS}. Depending on the efficiency of the amplification, dynamos generally require $B$-seeds stronger than $\sim$$10^{-22}$~G at the time of completed galaxy formation. It has also been claimed that this lower limit could be pushed down to $\sim$$10^{-30}$~G in spatially open, or in $\Lambda$-dominated FRW models~\cite{DLT}. The~size of the initial magnetic seed is also an issue, since it should not be smaller that $\sim$$100$~pc after the collapse of the protogalaxy, which implies a comoving scale of approximately $10$~Kpc before the collapse. Seeds~generated after inflation, during the radiation era for example, are typically too small in size because their coherence length can never exceed that of the causal horizon at the time of magnetogenesis. Inflation naturally achieves superhorizon correlations, so it can easily produce primordial fields of the required length. Nevertheless, in all the standard scenarios of inflationary magnetogenesis, $B$-fields decay adiabatically as soon as they cross outside the Hubble horizon. This~practically means that $B\propto a^{-2}$, with $a=a(t)$ representing the cosmological scale factor, essentially from the beginning of inflation until today. The result is astrophysically irrelevant magnetic fields today. In particular, the residual strength of a field with comoving (pre-collapse) size close to 10~Kpc today can be as low as $10^{-53}$~G (see~\cite{KKT,Wetal} and references therein). Having said that, the aforementioned numerical result assumes that the adiabatic magnetic decay persists on all scales after inflation. This is believed to reflect the high electrical conductivity of the post-inflationary universe, which in turn is thought to guarantee that magnetic fields remain frozen into the cosmic medium. The magnetic-flux freezing, however, is a causal (local) process, which cannot be achieved without the presence of electric currents. Therefore, applying the adiabatic decay-law on all scales, implicity assumes the existence of electric currents with superhorizon correlations, or that local causal physics can affect superhorizon perturbations. Both of these hypotheses, however, violate causality.

During inflation there are no electric currents and their formation starts once the universe enters its reheating phase. The process is causal, which implies that the coherence size of the newly formed currents never exceeds that of the horizon. After inflation, the latter coincides with the Hubble radius. Hence, the same causality arguments that confine the post-inflationary $B$-fields within the Hubble scale (see above), also forbid the electric currents from achieving super-Hubble correlations. Without such large-scale currents, it is no longer safe to employ the ideal magnetohydrodynamic (MHD) approximation to study the evolution of magnetic fields on superhorizon lengths. After all, the ideal-MHD limit is the result of causal microphysical processes, which have local range only and cannot dictate the evolution of $B$-fields with super-Hubble correlations without violating causality. Put another way, as long as they remain superhorizon-sized, $B$-fields remain immune to causal physics and they are only affected by the background expansion, just like any other inflation-generated perturbation. All these mean that, on scales larger than the horizon, the magnetic flux is not necessarily conserved and the adiabatic ($B\propto  a^{-2}$) decay-law is not a priori guaranteed. In fact, a straightforward calculation shows that, after the end of inflation, superhorizon-sized $B$-fields obey a power-law solution, where one of the modes drops slower than the adiabatic. This slowly decaying mode can dominate and thus dictate the magnetic evolution depending on the initial conditions. The latter are decided by the evolution of the field during the de Sitter phase and by the nature of the transition from inflation to reheating (as well as to the subsequent epochs of radiation and the dust). Following Israel's work on ``junction conditions'', we discuss two typical and complementary scenarios, illustrating how superhorizon-sized magnetic fields can be superadiabatically amplified after inflation.

The rate of the aforementioned slowly decaying magnetic mode depends on the equation of state of the matter that fills the universe at the time. Throughout the reheating phase, in particular, one~finds that $B\propto a^{-3/2}$, which slows down further to $B\propto a^{-1}$ in the radiation era, before returning to the $B\propto a^{-3/2}$-law during the subsequent dust epoch. Thus, as long as they remain outside the Hubble radius, magnetic fields are superadiabatically amplified all along their post-inflationary evolution. Once back inside the horizon, however, the electric currents take over and quickly freeze the $B$-fields into the highly conductive plasma, thus ``restoring'' their adiabatic decay-law. The time of the second horizon entry is crucial for the final magnetic strength and depends on the scale of the magnetic mode in question. Clearly, the larger the wavelength of the mode the longer its stays outside the Hubble radius, the longer its superadiabatic amplification and the stronger its residual magnitude. Assuming~a~magnetic seed with current comoving (pre-collapse) scale around 10~Kpc, for example, one can show that its present value is approximately $10^{-33}$~G. This is far stronger than the previously quoted value of $10^{-53}$~G. Further amplification is expected to occur during the collapse of the magnetised protogalactic cloud, which can bring the final strength of the field closer to, or even within, the galactic-dynamo range.

Although our analysis is primarily focused on the spatially flat FRW models, our conclusions and results extend naturally to their marginally closed and marginally open counterparts. This is intuitively plausible and it can also be shown analytically. Moreover, in marginally open Friedmann models, the superadiabatic amplification extends to the (physically unambiguous) subcurvature magnetic modes. Consequently, classical electromagnetism and conventional FRW cosmology can produce cosmological $B$-fields with residual strengths much larger than generally expected. Moreover, using the same arguments, one can either constraint or assist the non-conventional scenarios of primordial magnetogenesis, which amplify their fields during inflation and then allow them to decay adiabatically to the present. In particular, the more efficient the inflationary amplification, the stronger the constraint. A relatively mild amplification during the de Sitter phase, on the other hand, could produce $B$-fields of astrophysical relevance today. The main message, however, is that causality and the inferred absence of superhorizon-sized electric currents appear to make the post-inflationary evolution of large-scale $B$-fields a matter of initial conditions. These do not always guarantee the adiabatic decay-law, but also allow for the superadiabatic amplification of primordial magnetic fields on super-Hubble lengths. A~development that could put the question of cosmic magnetism under an entirely new perspective.

\section{The Question of Cosmic Magnetogenesis}\label{sQCM}
The scenarios of cosmic magnetogenesis are typically classified into early-time and late-time mechanisms, according to whether they operate before or after recombination. In this section we will briefly outline the main problems faced by the early-time mechanisms, which are distinguished further into inflationary and post-inflationary scenarios.

\subsection{The Scale Question}\label{ssScQ}
The main drawback of primordial magnetic fields generated after inflation is their scale. Typical~dynamos require seeds with coherence lengths no less than 100~pc by the time galaxy formation has been completed. This translates into a comoving scale of approximately 10~Kpc before the collapse of the proto-galactic cloud. Post-inflationary magnetic fields, however, are generally much smaller in size~\cite{KKT,Wetal}. The reason is causality, which always confines the correlation length of the generated $B$-field within that of the causal horizon (i.e.,~the Hubble radius). Put another way, given that no physical process propagates faster than the speed of light, all causally produced magnetic fields have sizes smaller than the Hubble length at the time of their creation\footnote{The same causality arguments that confine the coherence scale of the generated magnetic fields within the Hubble horizon, also restrict the correlation length of the newly formed electric currents (see~Section~\ref{ssCL-SME} below). Although the former constraint is a serious setback for most post-inflationary $B$-fields, the latter could prove a ``blessing in disguise'' for their inflationary counterparts (see Sections~\ref{sMFFFRWUs} and \ref{sRMF}).}. The latter is typically too small. For instance, assuming that the magnetic field is produced at the electroweak phase transition, its present size will be close to that of our solar system.

Theoretically, the scale problem can be solved, if there is an amount of MHD turbulence in the plasma and the initial $B$-seed is highly helical. In that case, magnetic helicity cascades inversely from smaller to larger scales, shifting magnetic energy to larger wavelengths and thus increasing the effective size of the original seed~\cite{BEO}-\cite{CHB}. Nevertheless, the present view is that the ``inverse cascade'' scenario is rather unlikely to deliver the desired results, unless the amount of primordial magnetic helicity is unrealistically large.

\subsection{The Strength Question}\label{ssStQ}
There is no scale issue whatsoever for magnetic fields generated during inflation. What the de Sitter phase does, is stretch subhorizon-sized quantum fluctuations in the Maxwell field to scales far larger than the Hubble radius, where they can be treated as classical electromagnetic fields. The main problem of inflationary magnetogenesis is the anticipated extreme weakness of the residual $B$-field, which is believed to have no astrophysical significance. Recall that galactic dynamos typically need magnetic seeds between $\sim$$10^{-22}$~G and $\sim$$10^{-12}$~G at the time the galaxy has been formed, although it might be possible to push the lower limit down to $\sim$$10^{-30}$ in open or in $\Lambda$-dominated FRW universes. This can happen because galaxies are older in the latter models, thus giving the dynamo more time to produce the observed $\mu$Gauss-order fields~\cite{DLT}.

The problem is that magnetic fields that have survived the de Sitter phase are largely expected to have strengths far below $10^{-30}$~G today. This has been attributed to the so-called adiabatic magnetic decay. The belief, in other words, that conventional $B$-fields decay as $B\propto a^{-2}$ ($a$ is the scale factor of the universe) at all times and on all scales. As a result, the typical strengths of inflationary magnetic fields quoted in the literature are below $10^{-50}$~G today. In particular, the residual magnitude of the $B$-field does not depend on the particulars of the adopted inflationary scenario and is given by (e.g.,~see~\cite{KKT,Wetal} and references therein)
\begin{equation}
B\simeq 10^{-57}\lambda_B^{-2} \hspace{2mm} {\rm G}\,,  \label{B*1}
\end{equation}
where $\lambda_B$ is the present (pre-collapse) scale of the magnetic mode in question (measured in Mpc). Setting $\lambda_B\simeq10$~Mpc, which is the minimum required for the dynamo to work (see Section~\ref{ssScQ} above), we find $B\simeq10^{-53}$ today. Therefore, unless classical Maxwellian electromagnetism or standard cosmology are abandoned, inflationary magnetic fields are astrophysically irrelevant. In the following sections we will demonstrate that this is not necessarily the case.

\section{Magnetic Fields in Flat FRW
Universes}\label{sMFFFRWUs}
Cosmological magnetic fields in spatially flat FRW universes are widely believed to decay adiabatically on all scales, during both their inflationary and post-inflationary life. Nevertheless, the adiabatic magnetic decay on superhorizon lengths has never been explicitly shown to hold, but its validity has been somehow heuristically extended from the sub-Hubble to the super-Hubble scales. Here, we will take another look at the evolution of large-scale $B$-fields after inflation.

\subsection{Causality and Large-Scale Magnetic 
Evolution}\label{ssCL-SME}
Consider an FRW spacetime, with Euclidean spatial geometry, permeated by weak electromagnetic perturbations. Then, introduce a group of observers with 4-velocity $u_a$ (so that $u_au^a=-1$). Relative to these observers, the electromagnetic tensor ($F_{ab}$) splits into an electric ($E_a$) and a magnetic ($B_a$) field as $F_{ab}=2u_{[a}E_{b]}+ \varepsilon_{abc}B^c$, with $\varepsilon_{abc}$ representing the 3-dimensional Levi-Civita tensor~\cite{TB}-\cite{BMT}. To linear order, the magnetic component of the Maxwell field obeys the wave-like formula~\cite{T1}
\begin{equation}
\ddot{B}_a+ 5H\dot{B}_a+ 3(1-w)H^2B_a- {\rm D}^2B_a= {\rm curl}\mathcal{J}_a\,,  \label{ddotBa}
\end{equation}
where $H=\dot{a}/a$ is the background Hubble parameter, $w=p/\rho$ is the barotropic index of the matter (with $\rho$ and $p$ representing its energy density and isotropic pressure respectively), ${\rm D}^2={\rm D}^a{\rm D}_a$ is the 3-dimensional covariant Laplacian operator and $\mathcal{J}_a$ the electric current (with $\mathcal{J}_au^a=0$). Also, ${\rm curl}\mathcal{J}_a=\varepsilon_{abc}{\rm D}^b\mathcal{J}^c$ by definition. Note that the above holds on a spatially flat FRW background (for the open and closed Friedmann models see Sections~\ref{ssMOFRWMs} and \ref{ssMCFRWMs} respectively). To simplify the mathematics let us introduce the rescaled magnetic field $\mathcal{B}_a=a^2B_a$ and use conformal, instead of proper, time ($\eta$ with $\dot{\eta}=1/a$). Then, expression (\ref{ddotBa}) reduces to the familiar compact form~\cite{KKT,Wetal}
\begin{equation}
\mathcal{B}^{\prime\prime}_a- a^2{\rm D}^2\mathcal{B}_a= a^2{\rm curl}\mathcal{J}_a\,,  \label{cBa''1}
\end{equation}
where the primes denote conformal-time derivatives. In addition to its compactness, the above expression is (formalistically) independent of the $w$-index, namely of the type of matter that fills the universe, provided the latter retains its barotropic nature.

During inflation the universe is believed to be a very poor electrical conductor. This means that there are no electric currents and during the de Sitter phase the right-hand side of (\ref{ddotBa}) vanishes identically. In other words, at the moment the inflation-produced magnetic fields exit the Hubble horizon they obey the wave-like equation
\begin{equation}
\mathcal{B}^{\prime\prime}_a- a^2{\rm D}^2\mathcal{B}_a= 0\,.  \label{cBa''2}
\end{equation}
Introducing the harmonic splitting $\mathcal{B}_a= \sum_n\mathcal{B}_{(n)}\mathcal{Q}_a^{(n)}$, with ${\rm D}_a\mathcal{B}_{(n)}=0=\mathcal{Q}_a^{\prime\,(n)}$ and ${\rm D}^2\mathcal{Q}_a^{(n)}=-(n/a)^2\mathcal{Q}_a^{(n)}$, the above recasts into
\begin{equation}
\mathcal{B}^{\prime\prime}_{(n)}+ n^2\mathcal{B}_{(n)}= 0\,, \label{cBn''}
\end{equation}
with $n>0$ representing the comoving wavenumber of the $n$-th magnetic mode. This differential equation accepts an oscillatory solution, which written for the actual magnetic field reads
\begin{equation}
a^2B_{(n)}= \mathcal{C}_1\cos(n\eta)+ \mathcal{C}_2\sin(n\eta)\,, \label{Bn}
\end{equation}
where $n\eta=\lambda_H/\lambda_n$. The latter ratio measures the physical size of the magnetic mode ($\lambda_n=a/n$) relative to the Hubble horizon ($\lambda_H=1/H$). Solution (\ref{Bn}) applies to inflationary magnetic fields as they cross the horizon during the de Sitter era. Once well outside the Hubble radius, namely on wavelengths with $\lambda_H/\lambda_n\ll1$ (i.e.,~for $n\eta\ll1$ in conformal-time terms), a simple Taylor expansion reduces the above to the power law
\begin{equation}
a^2B_{(n)}= \mathcal{C}_1+ \mathcal{C}_2n\eta\,,  \label{lsBn}
\end{equation}
with $a=a(\eta)$. The transition from oscillation to power-law growth at the Hubble threshold, as seen in solutions (\ref{Bn}) and (\ref{lsBn}), is nothing new to cosmological perturbation theory. It happens to linear density perturbations, for example, during the radiation epoch (e.g.,~see Section~4.4 in~\cite{P}). Physically, the change from oscillatory behaviour to power law at the Hubble length, simply reflects the fact that superhorizon-sized perturbations have not yet started to oscillate properly, because they have oscillation periods longer than the age of the universe at the time.\footnote{Recall that $\lambda_H=1/H\simeq t_u$ and $\lambda_n=t_n$, with $t_u$ and $t_n$ representing the age of the universe and the period of the magnetic-mode oscillation respectively. Then, on scales well beyond the Hubble radius (i.e.,~when $\lambda_n\gg\lambda_H$) we have $t_n\gg t_u$, which means that the oscillation has not yet reached its first wave-crest.}

Once outside the Hubble radius and as long as it stays there, the $B$-field remains causally disconnected and its evolution is only affected by the background expansion. Although the electrical conductivity of the universe grows after inflation and currents start to form, causality confines them inside the Hubble horizon. This ensures that there can never exist electric currents with superhorizon correlations and, in their absence, the ideal-MHD limit should not be applied to super-Hubble scales. Recall that it is the currents that eliminate the electric fields and freeze their magnetic counterparts into the matter.\footnote{The electrical properties of a medium are reflected in Ohm's law, which in its covariant form reads $\mathcal{J}_a=\varsigma E_a$, with $\varsigma$ representing the conductivity scalar~\cite{G,J}. Consequently, to eliminate a~superhorizon-sized electric field, requires the presence of currents coherent on the same scales. Given~that causality forbids the existence of such currents, the $E$-field will not vanish, unless it is fragmented into smaller (individually causally connected) parts. Nevertheless, even if we assume that the electric field has somehow been fragmented and eliminated by the local currents, its superhorizon-sized magnetic counterpart is ``unaware'' of that for as long as it remains causally disconnected.} Moreover, the process of magnetic-flux freezing is also causal and, as it is well known, \textit{causal physics can never affect superhorizon-sized perturbations}. This principle summarises the implications of causality for cosmology (e.g.,~see~\cite{RW}-\cite{Ba} for analogous quotes) and is at the root of the celebrated ``horizon problem''. Put another way, causality implies that the time required for the freezing-in information to travel the whole length of a super-Hubble $B$-field is longer than the age of the universe at the time. Therefore, the $B$-field cannot re-adjust itself to the new environment and freeze-in, until it has crossed back inside the horizon and come into full causal contact. Instead, as long as it remains outside the Hubble horizon, the magnetic field is immune to causal physics and retains only the ``memory'' of its distant past. This implies that the magnetic evolution is still governed by the long-wavelength limit (\ref{lsBn}) of the source-free  wave-equation (\ref{cBa''2}). In the following sections we will consider the implications of this claim.

It is worth noting that this is not the first time the aforementioned source-free approach is applied to the study of large-scale cosmological magnetic fields (e.g.,~see~\cite{TK}-\cite{T2}). Nevertheless, in~\cite{TK}-\cite{Ko} the role of causality and its implications were either assumed implicitly, or there was only a brief (passing) reference to them. Also, in~\cite{TK}-\cite{SS}, the background model was a spatially open Friedmann universe. The causality issue is discussed in certain detail also in~\cite{T2}, where the focus is on the non-conventional scenarios of cosmic magnetogenesis, and in~\cite{T3}, which looks specifically into the role of the initial conditions in conventional mechanisms of magnetic generation. In both of the aforementioned articles, the background universe is a spatially flat FRW spacetime. Here, we provide an extended discussion of the matter and of its potentially pivotal implications for cosmic magnetogenesis (conventional or not), in flat and in marginally curved Friedmann models.

\subsection{Large-Scale Superadiabatic Magnetic
Amplification}\label{ssL-SSMA}
Following the above arguments, large-scale (causally disconnected) magnetic fields evolve in line with the power-law solution (\ref{lsBn}), from the moment they cross outside the Hubble horizon during inflation until the time of their re-entry (in the radiation epoch, or later in the dust era). In what follows, we will focus our attention to the second mode on the right-hand side of (\ref{lsBn}), the presence of which implies that the adiabatic decay-law ($B\propto a^{-2}$) is not necessarily guaranteed on super-Hubble lengths. This mode is not a priori negligible, despite the fact that $n\eta\ll1$ on super-Hubble scales. Indeed, when the initial conditions are such that $\mathcal{C}_2\gg\mathcal{C}_1$, the aforementioned second mode can make a difference and it can lead to the superadiabatic amplification of large-scale magnetic fields (see Sections ~\ref{ssL-SSMA}--\ref{ssRICs} below). Clearly, as the universe expands, the conformal time increases and the product $n\eta$ will eventually become larger than unity. Physically this means that the $B$-field has re-entered the Hubble radius. Once back inside the horizon, solution (\ref{lsBn}) is no longer valid. There, causal physics take over and the electric currents can quickly freeze the magnetic field into the highly conductive cosmic medium. Then onwards, the ideal-MHD limits applies, the magnetic flux remains conserved and the $B$-field decays adiabatically (i.e.,~$B\propto a^{-2}$).

Before proceeding to examine the implications of solution (\ref{lsBn}) for the magnetic evolution after inflation, let us take a brief look at the inflationary phase first. Assuming exponential (de Sitter-type) expansion, we may set $a\propto-1/\eta$ with $\eta<0$. Then, after dropping the mode-index ($n$) for the economy of the presentation and then calculating the integration constants on the right-hand side of (\ref{lsBn}), the latter recasts into
\begin{equation}
B= \left(3B_0-\eta_0B^{\prime}_0\right) \left({a_0\over a}\right)^2- \left(2B_0-\eta_0B^{\prime}_0\right) \left({a_0\over a}\right)^3\,. \label{dSBn}
\end{equation}
Consequently, large-scale (conventional) magnetic fields on spatially flat FRW backgrounds decay adiabatically (i.e.,~$B\propto a^{-2}$) throughout the de Sitter phase of the expansion.

Let us now look at the evolution of superhorizon-sized cosmological $B$-fields after inflation. Once~again, after evaluating the two integration constants on the right-hand side of solution (\ref{lsBn}), the latter acquires the form shown below\footnote{Calculating the integration constants of (\ref{lsBn}) gives $\mathcal{C}_1=[B_0-\eta_0(2a_0H_0B_0+B_0^{\prime})]a_0^2$ and $\mathcal{C}_2=\eta_0(2a_0H_0B_0+B_0^{\prime})a_0^2/n\eta_0$. Given that $n\eta_0\ll1$ on super-Hubble scales, we deduce that $\mathcal{C}_2\gg\mathcal{C}_1$ (unless $2a_0H_0B_0+B_0^{\prime}=0$). This becomes clearer when the integration constants are evaluated in a~specific cosmic era. During reheating and dust, for example, $a\propto\eta^2$ and therefore $aH=a^{\prime}/a=2/\eta$. Then, $\mathcal{C}_1=-(3B_0+\eta_0B_0^{\prime})a_0^2$ and $\mathcal{C}_2=(4B_0+\eta_0B_0^{\prime})a_0^2/n\eta_0$, which guarantees that $\mathcal{C}_2\gg\mathcal{C}_1$ (unless $4B_0+\eta_0B_0^{\prime}=0$---see Equation~(\ref{rh-dB1}) in Section~\ref{ssERRD}). All these explain why one should not a priori discard the second mode of solution (\ref{lsBn}) before evaluating the integration constants first.}
\begin{equation}
B= \left[B_0-\eta_0\left(2a_0H_0B_0+B_0^{\prime}\right)\right] \left({a_0\over a}\right)^2+ \eta_0\left(2a_0H_0B_0+B_0^{\prime}\right) \left({a_0\over a}\right)^2\left({\eta\over\eta_0}\right)\,.  \label{lsB1}
\end{equation}
Note that we have used the relation $H=a^{\prime}/a^2$ for the Hubble parameter (recall that the primes indicate differentiation with respect to the conformal time). The above monitors the linear evolution of superhorizon-sized magnetic fields on spatially flat FRW backgrounds. We should also point out that the barotropic index of the matter is not necessarily constant but it can vary with time (i.e.,~$w=w(t)$). This means that solution (\ref{lsB1}) applies continuously throughout the lifetime of the universe, provided the cosmological expansion is entirely smooth and the matter can always be treated as a single barotropic medium. Under this proviso, expression (\ref{lsB1}) also monitors the magnetic evolution through the various cosmological transitions (e.g.,~the one leading from inflation to reheating).

The precise physics of the early transitions and the exact nature of the cosmic medium during those periods are still ambivalent. Nevertheless, the barotropic index of the matter is believed to maintain constant value during prolonged periods in the lifetime of the universe. As long as $w$ remains invariant, the cosmological scale factor and the conformal time are related by
\begin{equation}
a= a_0\left({\eta\over\eta_0}\right)^{2/1+3w}\,,  \label{sf-ct}
\end{equation}
where $w\neq-1/3$ and the zero suffix indicates a given initial time. On using the above, it is straightforward to show that $H=a^{\prime}/a^2=2/(1+3w)a\eta$ and then recast solution (\ref{lsB1}) into
\begin{equation}
B= -\left[\left({4\over1+3w}-1\right)B_0+\eta_0B^{\prime}_0\right] \left({a_0\over a}\right)^2+ \left({4B_0\over1+3w}+\eta_0B^{\prime}_0\right) \left({a_0\over a}\right)^{3(1-w)/2}\,.  \label{lsB2}
\end{equation}
The latter also monitors the linear evolution of superhorizon-sized $B$-fields on spatially flat FRW backgrounds filled with a single barotropic medium. In contrast to solution (\ref{lsB1}), however, here the barotropic index of the matter has been treated as a constant. Consequently, solution (\ref{lsB2}) does not apply continuously throughout the evolution of the universe, but only to periods during which $w=$\;constant\;$\neq-1/3$ (e.g.,~to the reheating and the radiation eras when $w=0$ and $w=1/3$ respectively). In other words, solution (\ref{lsB2}) is a special case of (\ref{lsB1}).

\subsection{The Epochs of Reheating, Radiation and
Dust}\label{ssERRD}
Looking at solutions (\ref{lsB1}) and (\ref{lsB2}), we immediately notice that the first of the two magnetic modes on their right-hand side always decays adiabatically. The rate of the second mode, however, is not a~priori fixed but depends on the equation of state of the cosmic medium. The latter also determines the relation between the cosmological scale factor and the conformal time. In particular, as long as $w=$~constant$\,>-1/3$ the second mode on the right-hand side of (\ref{lsB2}) decays at a rate slower than the adiabatic. The same behaviour can also be seen in solution (\ref{lsB1}). Therefore, when dealing with conventional matter, superhorizon sized magnetic fields on spatially flat FRW backgrounds are superadiabatically amplified. This, under the proviso that the initial conditions allow the second modes in (\ref{lsB1}) and (\ref{lsB2}) to survive and dominate.

With these in mind, let us take a closer look at the post-inflationary magnetic evolution. During~the reheating phase, as well as during the dust era later, $w=0$ , $a\propto\eta^2$ and $H=2/a\eta$. Then, solutions (\ref{lsB1}) and (\ref{lsB2}) reduce to
\begin{equation}
B= -\left(3B_0+\eta_0B^{\prime}_0\right) \left({a_0\over a}\right)^2+ \left(4B_0+\eta_0B^{\prime}_0\right) \left({a_0\over a}\right)^{3/2}\,.  \label{rh-dB1}
\end{equation}
Thus, as long as reheating lasts (as well as after equipartition) superhorizon-sized magnetic fields drop as $B\propto a^{-3/2}$, instead of following the standard adiabatic ($B\propto a^{-2}$) decay-law. During the intermediate epoch of radiation $w=1/3$, which means that $a\propto\eta$ and $H=1/a\eta$. Then, throughout that period solutions (\ref{lsB1}) and (\ref{lsB2}) take the form
\begin{equation}
B= -\left(B_0+\eta_0B^{\prime}_0\right) \left({a_0\over a}\right)^2+ \left(2B_0+\eta_0B^{\prime}_0\right) \left({a_0\over a}\right)\,,  \label{rB1}
\end{equation}
ensuring that large-scale magnetic fields drop as $B\propto a^{-1}$ when radiation dominates the energy density of the universe. Finally, let us also consider a phase of stiff-matter domination. In that case, $w=1$, $a\propto\eta^{1/2}$, $H=1/2a\eta$ and
\begin{equation}
B= -\eta_0B^{\prime}_0\left({a_0\over a}\right)^2+ \left(B_0+\eta_0B^{\prime}_0\right)\,,  \label{smB1}
\end{equation}
with the dominant mode remaining constant. Note that towards the end of inflation, when the inflaton rolls down the slope of its potential, the effective equation of state of the cosmic medium is that of stiff~matter.

In summary, after the end of the de Sitter phase, large-scale $B$-fields on spatially flat FRW backgrounds obey solutions which always contain modes with decay rates slower than the adiabatic. This happens without the need to break away from conventional electromagnetic theory, or to abandon standard physics and conventional cosmology. Whether these slowly-decaying magnetic modes can dominate over the adiabatic one depends on their associated coefficients. When the latter are of roughly the same order of magnitude, the slowly decaying modes quickly take over and dictate the subsequent evolution of the $B$-field. The initial conditions at the beginning of the post-inflationary epoch are therefore crucial.

\subsection{The Role of the Initial Conditions}\label{ssRICs}
The initial conditions of the post-inflationary magnetic evolution are decided by the field's behaviour in the de Sitter phase and by the nature of the transitions to the eras of reheating and radiation. Based on Israel's work on junction conditions~\cite{I}, we will discuss two typical and complementary initial-condition scenarios. Alternative approaches may also be possible.\vspace{6pt}

\textit{Scenario A:} Let us consider the typical scenario, where the background barotropic index undergoes an abrupt change from $w_*^-$ before the transition to $w_*^+$ afterwards (with $w_*^+\neq w_*^-$).\footnote{The $*$-suffix marks the moment the universe crosses from one epoch to the next. Also, the "$-$" and "$+$"~superscripts indicate the end of the era prior to the transition and the beginning of the next respectively.} Let us also assume that the matching spatial hypersurface is that of constant conformal time. This translates into a~``jump'' in the expansion rate of the background universe, namely in the Hubble parameter, on either side of the transit surface (i.e.,~$H_*^+\neq H_*^-$, or $[H_*]^+_-=H_*^+- H_*^-\neq0$). The latter implies a discontinuity in the extrinsic curvature of the matching hypersurface, which requires the presence of a ``thin shell'' there with finite energy-momentum tensor. Practically speaking, we assume that the width of the shell is too small compared to the scales of interest. In that case, the aforementioned shell can be replaced by a spacelike hypersurface. Discontinuities of this nature can be used to bypass the (as yet ambivalent) details of early cosmological transitions, like the one leading from inflation to reheating (e.g.,~see~\cite{CW} and references therein).

Conventional scenarios of inflationary magnetogenesis demand that the magnetic field decays adiabatically throughout the de Sitter regime (i.e.,~$B\propto a^{-2}$). On the other hand, typical non-conventional mechanisms of primordial magnetic generation amplify their $B$-fields superadiabatically during inflation (i.e.,~$B\propto a^{-m}$ with $0\leq m<2$)~\cite{KKT,S}. With these in mind, let us assume that all along the de Sitter phase the magnetic field obeys the power law
\begin{equation}
B= B_0\left({a_0\over a}\right)^m= B_0\left({\eta\over\eta_0}\right)^m\,,  \label{dSB1}
\end{equation}
where $0\leq m\leq2$ and the zero suffix indicates the beginning of the exponential expansion. Note that the second equality reflects the fact that $a\propto-1/\eta$, with $\eta<0$, during de Sitter-type inflation (i.e.,~for $w=-1$ -- see Equation~({\ref{sf-ct}) in Section~\ref{ssERRD}). Differentiating (\ref{dSB1}) with respect to the conformal time gives $B^{\prime}=mB/\eta$, which ensures that
\begin{equation}
\eta_*^-B_*^{\prime\;-}= mB_*^-\,. \label{dSfcon1}
\end{equation}
at the end of inflation proper.

Once into reheating, the barotropic index changes from $w_*^-=-1$ to $w_*^+=0$. Then, according to solution (\ref{lsB2}), throughout reheating superhorizon-sized magnetic fields evolve as
\begin{equation}
B= -\left(3B_*^++\eta_*^+B^{\prime\;+}_*\right) \left({a_*^+\over a}\right)^2+ \left(4B_*^++\eta_*^+B^{\prime\;+}_*\right) \left({a_*^+\over a}\right)^{3/2}\,,  \label{rhB+1}
\end{equation}
with $a\geq a_*^+$ (see also Equation~(\ref{rh-dB1}) in Section~\ref{ssERRD}). When the transit hypersurface is that of constant conformal time, we may set $\eta_*^+=-\eta_*^-$ (recall that $\eta_*^+>0$ and $\eta_*^-<0$).~\footnote{Setting $\eta_*^+=-\eta_*^-$ on either side of the transit hypersurface does not constitute a real discontinuity in the conformal time. The~jump ($[\eta_*]^+_-=2\eta_*^+$) is only an apparent one, since it can be removed by replacing $\eta$ with the variable $x=|\eta|$.} This implies a ``jump'' in the expansion rate of the background universe and a discontinuity in its extrinsic curvature of the matching hypersurface, which can be compensated by the presence of a thin layer there~\cite{CW}. Consequently, assuming that there is no magnetic discontinuity at the linear level, namely that $B_*^+=B_*^-$ and $B_*^{\prime\;+}=B_*^{\prime\;-}$, constraint (\ref{dSfcon1}) translates into
\begin{equation}
\eta_*^+ B_*^{\prime\;+}= -mB_*^+\,.  \label{rhicon+1}
\end{equation}
The above sets the initial conditions for the evolution of the $B$-field during reheating and combines with solution (\ref{rhB+1}) to give
\begin{equation}
B= -(3-m)B_*^+\left({a_*^+\over a}\right)^2+ (4-m)B_*^+\left({a_*^+\over a}\right)^{3/2}\,,  \label{rhB+2}
\end{equation}
where $a\geq a_*^+$.\footnote{We can obtain the evolution law (\ref{rhB+2}) starting from solution (\ref{lsB1}) as well. This requires calculating the jump in the value of the Hubble parameter caused by the abrupt change of the barotropic index on the matching hypersurface, which is that of constant conformal time. In order to do that recall first that $H_*^-=-1/a_*^-\eta_*^-$ at the end of the de Sitter regime and $H_*^+=2/a_*^+\eta_*^+$ at the start of reheating. Then, using conditions (\ref{dSfcon1}) and (\ref{rhicon+1}), while setting $a_*^+=a_*^-$, $\eta_*^+=-\eta_*^-$, $B_*^+=B_*^-$ and $B_*^{\prime\;+}=B_*^{\prime\;^-}$, solution (\ref{lsB1}) reduces to Equation~(\ref{rhB+2}).} Therefore, as long as $m\neq4$, the dominant magnetic mode of (\ref{rhB+2}) drops as $B\propto a^{-3/2}$ and the $B$-field is superadiabatically amplified throughout reheating. We remind the reader that almost all the scenarios of inflationary magnetogenesis assume that $0\leq m\leq2$.

Let us look at the magnetic evolution in the subsequent epochs of radiation and dust. Following~solution (\ref{rhB+2}) and keeping in mind that $a\propto\eta^2$ during reheating (see Equation~(\ref{sf-ct})), we deduce that $B\propto\eta^{-3}$ throughout that period. Then,
\begin{equation}
\eta_*^-B_*^{\prime\;-}= -3B_*^-\,,  \label{rhfcon}
\end{equation}
just before the transition to the radiation era. Once there, the barotropic index of the background matter changes from $w_*^-=0$ to $w_*^+=1/3$ and solution (\ref{lsB2}) reads
\begin{equation}
B= -\left(B_*^++\eta_*^+B_*^{\prime\;+}\right) \left({a_*^+\over a}\right)^2+ \left(2B_*^++\eta_*^+B_*^{\prime\;+}\right) \left({a_*^+\over a}\right)\,,  \label{rB+1}
\end{equation}
with $a\geq a_*^+$ (see also Equation~(\ref{rB1}) in Section~\ref{ssERRD}). As before, suppose that the matching hypersurface is that of constant conformal time and assume that the magnetic evolution through the transition is smooth. Then, demanding $\eta_*^+=\eta_*^-$, $B_*^+=B_*^-$ and $B_*^{\prime\;^+}=B_*^{\prime\;^-}$, constraint (\ref{rhfcon}) recasts into
\begin{equation}
\eta_*^+B_*^{\prime\;+}= -3B_*^+\  \label{ricon}
\end{equation}
and sets the initial conditions for the magnetic evolution in the radiation era. Substituting the above into the right-hand side of (\ref{rB+1}), we arrive at
\begin{equation}
B= 2B_*^+\left({a_*^+\over a}\right)^2- B_*^+\left({a_*^+\over a}\right)\,,  \label{rB+2}
\end{equation}
where $a\geq a_*^+$. Consequently, superhorizon-sized magnetic fields are superadiabatically amplified (i.e.,~$B\propto a^{-1}$) all along the radiation epoch as well.

Similarly, we find that $\eta_*^-B_*^{\prime\;-}= -B_*^-$ prior to the equilibrium time, since $a\propto\eta$ when $w=1/3$ (see Equation~(\ref{sf-ct}) in Section~\ref{ssERRD}). At the time of matter-radiation equality the background barotropic index changes from $w_*^-=1/3$ to $w_*^+=0$. Then, when the matching hypersurface is that of constant conformal time and the $B$-field evolves smoothly through the transit, we have
\begin{equation}
\eta_*^+B_*^{\prime\;+}= -B_*^+\,,  \label{dicon}
\end{equation}
at the start of matter domination. Finally, setting $w=w_*^+=0$ into the right-hand side of solution (\ref{lsB2}) and using the above initial conditions, leads to
\begin{equation}
B= -2B_*^+\left({a_*^+\over a}\right)^2+ 3B_*^+\left({a_*^+\over a}\right)^{2/3}\,,  \label{dB+1}
\end{equation}
with $a\geq a_*^+$. Therefore, as long as the magnetic field remains outside the Hubble radius, $B\propto a^{-3/2}$ and its superadiabatic amplification continues into the dust era as well. Moreover, the effect is independent of the magnetic evolution during the de Sitter phase. On whether, in particular, the $B$-field depleted adiabatically throughout inflation or not (provided $B\propto a^{-m}$, with $m\neq4$ at the time---see solution~(\ref{rhB+2})~above).
\vspace{6pt}

\textit{Scenario B:} Suppose that the background equation of state undergoes an abrupt change, as the universe crosses from one epoch to the next, but this time do not allow for a thin shell on the transition hypersurface. Then, there can be no discontinuity in the extrinsic curvature of the matching surface. When dealing with a Friedmann universe, this means no jump in the value of the background Hubble parameter there (i.e.,~$[H_*]^+_-= H_*^+-H_*^-=0$). In such a case, the transit hypersurface is that of constant energy density, though not necessarily of constant conformal time. Discontinuities of this nature can also be used to cope with early universe transitions.

In line with the literature on inflationary magnetogenesis and with Scenario~A before, let us assume that large-scale primordial magnetic fields obey the power law
\begin{equation}
B= B_0\left({a_0\over a}\right)^m\,,  \label{mB}
\end{equation}
during the de Sitter phase. Again, the zero suffix indicates the onset of the exponential expansion and $0\leq m\leq2$. Differentiating the above with respect to the conformal time, guarantees that
\begin{equation}
B_*^{\prime\;-}= -ma_*^-H_*^-B_*^-\,,  \label{icon}
\end{equation}
at the end of inflation proper. Recalling that $[a_*]^+_-=0=[H_*]^+_-$ on the background matching surface and then setting $[B_*]^+_-=0=[B_*^{\prime}]^+_-$ at the linear level, constraint (\ref{icon}) translates into
\begin{equation}
B_*^{\prime\;+}= -ma_*^+H_*^+B_*^+\,,  \label{trscon}
\end{equation}
at the start of reheating and sets the initial conditions for the subsequent evolution of the $B$-field.

Following solution (\ref{lsB1}), throughout the reheating phase (when $a\propto\eta^2$), superhorizon-sized magnetic fields are monitored by
\begin{equation}
B= \left[B_*^+-\eta_*^+\left(2a_*^+H_*^+B_*^+ +B_*^{\prime\;+}\right)\right] \left({a_*^+\over a}\right)^2+ \eta_*^+\left(2a_*^+H_*^+B_*^++B_*^{\prime\;+}\right) \left({a_*^+\over a}\right)^{3/2}\,,  \label{lsB+1}
\end{equation}
with $a\geq a_*^+$. Inserting condition (\ref{trscon}) into the right-hand side of the above and keeping in mind that $H=2/a\eta$ during reheating, we obtain
\begin{equation}
B= -(3-2m)B_*^+ \left({a_*^+\over a}\right)^2+ 2(2-m)B_*^+\left({a_*^+\over a}\right)^{3/2}\,,  \label{lsB+2}
\end{equation}
where $a\geq a_*^+$. When $m=2$ the second term on the right-hand side vanishes, leaving the adiabatic (i.e.,~$B\propto a^{-2}$) mode only. For $m\neq2$, however, the second mode of solution (\ref{lsB+2}) survives and the magnetic decay-rate slows down to $B\propto a^{-3/2}$. A straightforward calculation confirms that this pattern is repeated at the subsequent transitions to the radiation and the dust eras. Consequently, in the absence of thin shells on the transition hypersurfaces, only magnetic fields that decay adiabatically during a certain cosmological epoch will continue to do so for their subsequent evolution. When there is no adiabatic decay prior to the transit, the $B$-field is superadiabatically amplified after the transition (provided $w_*^+>-1/3$).

The implications of scenario~B for inflationary magnetogenesis are fairly straightforward to deduce. Primordial magnetic fields that happen to decay adiabatically throughout inflation will continue to do so for the rest of the lifetime. This is essentially the ``standard'' conventional scenario of primordial magnetogenesis, which produces $B$-fields with astrophysically irrelevant residual strengths. However, large-scale magnetic fields that did not obey the $B\propto a^{-2}$ law during the de Sitter phase will experience superadiabatic amplification after the end of inflation. This result can affect the non-conventional scenarios of primordial magnetogenesis that superadiabatically amplify their $B$-fields during inflation (see~Section~\ref{ssNCSs} below).

\section{The Residual Magnetic Field}\label{sRMF}
Following our discussion so far, depending on the initial conditions, conventional large-scale $B$-fields can be superadiabatically amplified throughout the post-inflationary evolution of a flat FRW universe. Next, we will estimate the residual strength of such fields.

\subsection{The Time of Second Horizon Crossing}\label{ssTSHC}
To begin with, recall that imposing the adiabatic decay-law at all times and on all scales has lead to magnetic fields of approximately $10^{-53}$~G today, when their current comoving size is close to 10~Kpc (see Equation~(\ref{B*1}) in Section~\ref{ssStQ}). Also, during the de Sitter regime, superhorizon-sized magnetic fields decay adiabatically as expected (see solution ({\ref{dSBn}) in Section~\ref{ssL-SSMA}). The situation changes after inflation, when the magnetic decay-rate slows down (see solutions (\ref{rh-dB1}) and (\ref{rB1}) in Section~\ref{ssERRD}). Throughout~reheating, in particular, we have $B\propto a^{-2/3}$. This slows down further (to $B\propto a^{-1}$) in the radiation era, before returning to the $B\propto a^{-2/3}$ law during the subsequent dust epoch.~\footnote{At the end of inflation the scalar field rolls down the slope of the potential to its minimum. Then, the effective equation of state of the inflaton field ($\phi$) is that of stiff matter, with $p_{\phi}\simeq \rho_{\phi}\simeq \dot{\phi}/2$. During these final stages we have $a\propto\sqrt{\eta}$, with $\eta>0$, which substituted into Equation~(\ref{Bn}) leads to $B=C_3(a_0/a)+C_4$ (see also solution (\ref{smB1})). In other words, for the brief period between the de Sitter phase and reheating, the dominant magnetic mode of (\ref{Bn}) remains constant. We are not going to consider the implications of the aforementioned epoch here.} As a result, the residual magnetic strength can be considerably larger than expected. The overall amplification depends on the scale of the magnetic mode in question, which determines the time of horizon entry. Recall that once inside the Hubble radius the adiabatic decay is restored. This occurs because on subhorizon scales the electric currents take over, eliminate the electric fields and freeze their magnetic counterparts into the highly conductive medium. Put another way, we can apply the ideal-MHD limit only after the second horizon crossing. Then onwards, the magnetic flux remains conserved and the $B$-field decays adiabatically (at the linear perturbative level).

Suppose that the current comoving scale of the magnetic seed is $\lambda_B\simeq10$~Kpc, which is the minimum required for the dynamo to work. Fields of this size have $(\lambda_H/\lambda_B)_{\dag}\simeq3\times10^5$, where the $\dag$-suffix denotes the present, assuming that $\lambda_H\simeq3\times10^3$~Mpc is the Hubble radius today. Given~that $\lambda_H\propto t$ and $\lambda_B\propto a$, we deduce that $\lambda_H/\lambda_B\propto a^{1/2}$ during the dust era (when $t\propto a^{3/2}$) and $\lambda_H/\lambda_B\propto a$ throughout the preceding radiation epoch (when $t\propto a^2$). Putting these together, one finds that scales close to 10~Kpc today entered the horizon at $a_{HC}\simeq1/3\times10^{-3}a_{EQ}$. The latter translates into $T_{HC}\simeq3\times10^{-6}$~GeV, since $T\propto a^{-1}$ at all times and $T_{EQ}\simeq10^{-9}$~GeV. Until then, the $B$-field was lying outside the Hubble radius and it was superadiabatically amplified.

\subsection{The Final Magnetic Strength}\label{ssFMS}
As mentioned above, magnetic fields decay adiabatically during the de Sitter phase and once they are back inside the horizon after inflation. Therefore, the superadiabatic amplification occurs from the end of inflation proper until the second horizon crossing. Suppose that $\rho_B=B^2$ is the magnetic energy density and $\rho$ that of the dominant matter component. Then, at the end of the de Sitter regime we have
\begin{equation}
\left({\rho_B\over\rho}\right)_{DS}\simeq 10^{-94} \left({M\over10^{17}}\right)^{4/3} \left({T_{RH}\over10^{10}}\right)^{-4/3} \lambda_B^{-4}\,.  \label{dSB}
\end{equation}
Note that $M$ is the scale of inflation, $T_{RH}$ is the reheat temperature (both measured in GeV) and $\lambda_B$ is the current physical scale (measured in Mpc) of the magnetic mode in question. During reheating, $\rho_B\propto a^{-3}$ and $\rho\propto a^{-3}$ as well. Therefore, throughout this phase, the dimensionless ratio $\rho_B/\rho$ remains unchanged, which means that $(\rho_B/\rho)_{RH}\simeq(\rho_B/\rho)_{DS}$. Once into the radiation era, however, $\rho_B\propto a^{-2}$ and $\rho\simeq\rho_{\gamma}\propto a^{-4}$, with $\rho_{\gamma}$ representing the energy density of the radiative component. Hence, for a~magnetic mode that crosses inside the Hubble horizon before equipartition,
\begin{equation}
\left({\rho_B\over\rho_{\gamma}}\right)_{HC}\simeq \left({\rho_B\over\rho}\right)_{RH} \left({T_{RH}\over T_{HC}}\right)^2\simeq 10^{-94} \left({M\over10^{17}}\right)^{4/3} \left({T_{RH}\over10^{10}}\right)^{-4/3}\left({T_{RH}\over T_{HC}}\right)^2 \lambda_B^{-4}\,.  \label{HCB}
\end{equation}

After horizon crossing $\rho_B$, $\rho_{\gamma}\propto a^{-4}$, ensuring that their ratio remains constant until today. In other words,
\begin{equation}
\left({\rho_B\over\rho_{\gamma}}\right)_{\dag}\simeq 10^{-94} \left({M\over10^{17}}\right)^{4/3} \left({T_{RH}\over10^{10}}\right)^{-4/3}\left({T_{RH}\over T_{HC}}\right)^2 \lambda_B^{-4}\,,  \label{fB1}
\end{equation}
today (recall that the $\dag$-suffix corresponds to the present). As we have seen in the previous section, magnetic fields with current comoving size close to 10~Kpc, re-enter the horizon at $T_{HC}\simeq3\times10^{-6}$~GeV. Substituting this value into the right-hand side of Equation~(\ref{fB1}) and recalling that $(\rho_{\gamma})_{\dag}\simeq10^{-51}~{\rm GeV}^4$, gives
\begin{equation}
B_{\dag}\simeq 10^{-33}\left({M\over10^{17}}\right)^{2/3} \left({T_{RH}\over10^{10}}\right)^{1/3} \hspace{2mm} {\rm G}\,.  \label{fB2}
\end{equation}
Therefore, when $M\simeq10^{17}$~GeV and $T_{RH}\simeq10^{10}$~GeV, the present magnitude of a cosmological magnetic field with current physical size around 10~Kpc is close to $10^{-33}$~G, instead of $10^{-53}$~G. In other words, by simply appealing to causality, one can increase the final strength of conventional inflationary magnetic seeds by roughly 20 orders of magnitude.

\section{Implications for Cosmic Magnetogenesis}\label{sICM}
Our results solely affect superhorizon-sized magnetic fields. This means that they do not interfere at all with the mechanisms of post-inflationary magnetogenesis, which produce subhorizon-sized $B$-fields only (for the aforementioned causal reasons). There are potentially pivotal consequences, however, for the inflationary scenarios, both the non-conventional and the conventional (see also~\cite{T2} and~\cite{T3} respectively).

\subsection{Conventional Scenarios}\label{ssCSs}
Conventional inflationary magnetic fields decay adiabatically during the de Sitter phase, but deplete at a slower pace after inflation. Here, this happens within scenario~A (see Section~\ref{ssRICs} earlier). In~general, any scenario that allows the second magnetic mode on the right-hand side of solution (\ref{rh-dB1}) to survive at the start of reheating will lead to the same result. Then, the residual comoving (pre-collapse) magnitude of a $B$-field (with physical scale close to 10~Kpc today) will be approximately $10^{-33}$~G. The~magnetic strengths required for the dynamo to work are estimated at the time of completed galaxy formation (see Section~\ref{ssStQ} earlier). The magnitude quoted above is comoving, which means that it does not include the magnetic amplification that occurs during the collapse of the proto-galactic cloud. Assuming an idealistic spherically symmetric collapse, we may add up to four orders of magnitude to the comoving magnetic strength. Adopting the more realistic scenario of anisotropic protogalactic collapse leads to further increase by one or two orders of magnitude~\cite{DBL1}-\cite{BrMT}. All these can bring the final magnetic strength close to $10^{-27}$~G by the time the galaxy is formed. This is stronger than $10^{-30}$~G, which is the minimum magnetic strength quoted in the literature as capable of seeding the dynamo~\cite{DLT}. Hence, astrophysically relevant magnetic fields are theoretically possible without violating conventional electromagnetism or abandoning standard cosmology.

Additional magnetic amplification may be possible as well. The literature contains mechanisms that could enhance cosmological $B$-fields during both the earlier and the later stages of their evolution. Turbulent motions, for example, can increase the final magnitude of the field, once the latter is well inside the Hubble horizon. Here, we would like to draw the reader's attention to an alternative possibility, which is directly related to our discussion. In line with solution (\ref{lsB2}), the stiffer the equation of state of the cosmic medium, the slower the magnetic decay and the stronger its superadiabatic amplification. In fact, for stiff-matter (i.e.,~at the $p=\rho$ and $w=1$ limit) we obtain
\begin{equation}
B= {\rm constant}\,,  \label{smB2}
\end{equation}
for the dominant magnetic mode (see also solutions (\ref{lsB2}) and ({\ref{smB1})). Given that the energy density of a stiff medium drops as $\rho\propto a^{-6}$, we deduce that $\rho_B/\rho\propto a^6$ as long as $p=\rho$ is the equation of state of the dominant matter component. This implies that a very brief stiff-matter epoch before the start of the radiation era could lead to a substantial magnetic amplification without necessarily affecting the observational constraints.~\footnote{A period of stiff-matter dominance, prior to the radiation era, was originally proposed by Zeldovich~\cite{Z}. Provided this epoch was sufficiently brief, it could have left fundamental physical processes, like primordial nucleosynthesis (with $T_{NS}\simeq1$~MeV), unaffected. The~possibility that a phase of stiff-matter domination could assist the survival of inflationary magnetic fields has also been raised in~\cite{FJS}.} In terms of temperature, the radiation epoch typically spans from $T_{RH}\simeq10^{10}$~GeV up to $T_{eq}\simeq10^{-9}$~GeV. Therefore, if the universe is dominated by stiff matter between, say, $T_{RH}\simeq10^{10}$~GeV and $T=T_{SM}\simeq10^7$~GeV, the residual magnetic strength will increase from $10^{-27}$~G to $10^{-21}$~G (see~\cite{T3} for the details), which lies within the typical galactic-dynamo requirements. A longer phase of stiff-matter domination, say up to $T_{SM}\simeq10^4$~GeV, will boost the final magnitude of the $B$-field close to $10^{-15}$~G, that is very close to the recently reported magnetic strengths in empty intergalactic space~\cite{NV}-\cite{CBF}. We also note that during the final stages of inflation, when the inflaton rolls down its potential, the effective equation of state is that of stiff matter. Thus, in principle at least, one might be able to take advantage of this brief period to further enhance the magnetic field's strength.

\subsection{Non-Conventional Scenarios}\label{ssNCSs}
The vast majority of the inflationary magnetogenesis mechanisms operate outside conventional electromagnetic theory, or introduce some other kind of new physics. There is a very long list of non-conventional scenarios and for this reason we direct the reader to~\cite{KKT,Wetal} for recent reviews and specific references, while a relatively brief discussion can be found in~\cite{Su}. In most of the proposed mechanisms the $B$-field is superadiabatically amplified (i.e.,~$B\propto a^{-m}$, with \mbox{$0<m<2$}) during the de Sitter phase. After that, standard electromagnetism is usually restored and the final magnetic strength is estimated by assuming that $B$-fields decay adiabatically until today. This assumption does not a priori hold, however, given that all the astrophysically relevant modes remain outside the horizon at least until late into the radiation era. On these scales, the aforementioned magnetic fields are superadiabatically amplified throughout their post-inflationary evolution within both of our initial-condition scenarios (see Section~\ref{ssRICs} earlier). Therefore, residual magnitudes based on the adiabatic-decay law after inflation need to be revised. As we will argue next, the revision will affect (to~a~larger or lesser degree) essentially all the mechanisms of primordial magnetogenesis that amplify their $B$-fields during inflation (see also~\cite{T2} for an extensive discussion).

Scenarios of inflationary magnetic amplification are often susceptible to backreaction problems. In other words, the Maxwell field can get strong enough to start interfering with the background kinematics. Even when there are no backreaction issues, however, there might be problems with the observational constrains. The large-scale magnetic fields observed in galaxies and in galactic clusters, for example, are close to $10^{-6}$~G and $10^{-7}$~G respectively. Also, the results of primordial nucleosynthesis and the high isotropy of the cosmic microwave background (CMB), seem to exclude $B$-fields with current strengths larger than $\sim$$10^{-7}$~G and $\sim$$10^{-9}$~G respectively~\cite{KKT,Wetal}. None of the aforementioned non-conventional scenarios of primordial magnetogenesis violates the above constraints, but only after assuming that the adiabatic decay-law holds from the end of inflation until today. When the $B$-field remains superadiabatically amplified throughout its entire post-inflationary evolution, however, one should probably revise the residual magnetic strengths and check whether they comply or not with the observations. In what follows we will consider two characteristic alternative scenarios to illustrate our argument. The first will allow for a rather strong superadiabatic amplification during the de Sitter regime (e.g.,~$B\propto a^{-m}$, with $0<m<1$), while in the second the amplification will be relatively mild (e.g.,~$B\propto a^{-m}$, with $1<m<2$).

Suppose that $B\propto a^{-1/2}$ throughout inflation, which implies relatively strong amplification during that period. Then, at the end of the de Sitter phase, the relative magnetic strength will be given \mbox{by the ratio}
\begin{equation}
\left({\rho_B\over\rho}\right)_{DS}\simeq 10^{-30} \left({M\over10^{17}}\right)^{10/3} \left({T_{RH}\over10^{10}}\right)^{-1/3} \lambda_B^{-1}\,,  \label{sinfamp1}
\end{equation}
where $M$ and $T_{RH}$ are the energy scale and the reheat temperature of the inflationary model respectively (both measured in GeV), while $\lambda_B$ is the current comoving scale of the field (in Mpc). For a magnetic mode that crosses inside the horizon at recombination we may set $\lambda_B\simeq3\times10^{3/2}$~Mpc at present. Ignoring reheating for simplicity, the above magnetic mode is superadiabatically amplified during the radiation era and for the brief period between equipartition and decoupling. In that case we have
\begin{equation}
B_{\dag}\simeq 10^{-2} \left({M\over10^{17}}\right)^{5/3} \left({T_{RH}\over10^{10}}\right)^{5/6} \hspace{2mm} {\rm G}\,,  \label{sinfamp2}
\end{equation}
today. For typical values of the inflationary parameters, for instance when $M\sim10^{17}$~GeV and $T_{RH}\sim10^{10}$~GeV, the above gives $B_{\dag}\gg10^{-9}$~G, in violation of the CMB constraints. Therefore, causality and the resulting absence of superhorizon-sized electric currents can essentially rule out a~host of primordial magnetogenesis mechanisms.

The situation changes drastically when the inflationary amplification of the $B$-field is relatively weak. For instance, let us assume that $B\propto a^{-3/2}$ throughout the de Sitter phase. Then, proceeding as before we find
\begin{equation}
\left({\rho_B\over\rho}\right)_{DS}\simeq 10^{-73} \left({M\over10^{17}}\right)^2 \left({T_{RH}\over10^{10}}\right)^{-1} \lambda_B^{-3}  \label{winfamp1}
\end{equation}
and subsequently
\begin{equation}
B_{\dag}\simeq 10^{-25} \left({M\over10^{17}}\right) \left({T_{RH}\over10^{10}}\right)^{1/2} \hspace{2mm} {\rm G}\,,  \label{winfamp2}
\end{equation}
for a magnetic mode that crossed the horizon around decoupling. Fields with the above (comoving) strength today are too weak to affect the CMB isotropy but strong enough to seed the galactic dynamo. Recall that a comoving magnitude of approximately $10^{-25}$~G can increase to roughly $\sim$$10^{-19}$~G by the time the galaxy is formed. So, in this case, the absence of large-scale electric currents and the resulting superadiabatic magnetic amplification on super-Hubble lengths appears to assist the associated scenarios of cosmic magnetogenesis, thus making them more promising candidates.

Overall, mechanisms of primordial magnetic generation leading to a substantial (superadiabatic-type) amplification of the $B$-field during inflation are likely to be in conflict with the observations. On the other hand, scenarios that achieve relatively mild enhancement during the de Sitter regime can produce magnetic seeds of real astrophysical relevance. For example, in~\cite{KSW} the authors discuss two (non-conventional) mechanisms of primordial magnetogenesis. One achieves strong magnetic enhancement during inflation, producing a $B$-field of approximately $10^{46}$~G on all scales by the end of the de Sitter expansion. In the other case, the amplification is mild and it only manages a~magnetic field around $10^{22}$~G on lengths close to $1$~Mpc (the scale has been redshifted to the present). The former field is too strong and triggers the aforementioned backreaction problems, while in our scenario its current magnitude violates all the available observational constrains. The latter magnetic field, however, has no such problems and, according to our scenario, its residual strength is capable of seeding the galactic dynamo (see~\cite{T2} for the details). All these suggest that the current limits put on inflationary magnetogenesis can be relaxed considerably. In particular, inflation-produced $B$-fields that are stronger than $\sim$$10^{17}$~G by the end of the de Sitter phase should be capable of seeding the galactic dynamo today (see~\cite{T2} for further discussion and more numerical results).

\section{The Case of Nearly Flat FRW Universes}\label{sCNFFRWUs}
Although we have so far confined our analysis to Friedmann universes with Euclidean spatial geometry, the same results also apply to FRW models with nearly flat spacelike hypersurfaces. \mbox{To a large} extent this may be intuitively obvious, but it can be shown analytically as well.

\subsection{Marginally Open FRW Models}\label{ssMOFRWMs}
We begin by recalling that, at the ideal-MHD limit, magnetic fields decay adiabatically irrespective of the background spatial curvature. Thus, in the presence of highly conductive electric currents, $B\propto a^{-2}$ at all times and in all three FRW spacetimes. When there are no currents, however, the magnetic evolution also depends on the geometry of the background universe. Throughout inflation, for example, or on superhorizon scales after the end of the accelerated expansion phase, the magnetic field obeys the wave-like equation~\cite{T1}
\begin{equation}
\mathcal{B}^{\prime\prime}_{(n)}+ \left(n^2+2K\right)\mathcal{B}_{(n)}= 0\,,  \label{pm1cBn}
\end{equation}
where $K=0,\pm1$ is the 3-curvature index.~\footnote{The presence of the spatial-curvature term in the magnetic wave equation can be seen as a reflection of the fact that Friedmannian spacetimes with nonzero 3-curvature are only locally conformal to the Minkowski space. Global conformal flatness applies only to FRW models with Euclidean spatial hypersurfaces.} In Friedmann models with negative spatial curvature (i.e.,~for $K=-1$), the above takes the form
\begin{equation}
\mathcal{B}^{\prime\prime}_{(n)}+ \left(n^2-2\right)\mathcal{B}_{(n)}= 0\,,  \label{-1cBn}
\end{equation}
with the comoving eigenvalue being positive and continuous (i.e.,~$n>0$). Eq.~(\ref{-1cBn}) accepts two qualitatively different families of solutions, depending on the range of the associated eigenvalues. When $n^2<2$, in particular, we find hyperbolic behaviour with
\begin{equation}
\mathcal{B}_{(n)}= \mathcal{C}_1\cosh\left[\left(\sqrt{2-n^2}\,\right)\eta\right]+ \mathcal{C}_2\sinh\left[\left(\sqrt{2-n^2}\,\right)\eta\right]\,.  \label{hcB}
\end{equation}
On the other hand, as we move to smaller scales (those with $n^2>2$), we recover the more familiar oscillatory evolution,
\begin{equation}
\mathcal{B}_{(n)}= \mathcal{C}_3\cos\left[\left(\sqrt{n^2-2}\right)\eta\right]+ \mathcal{C}_4\sin\left[\left(\sqrt{n^2-2}\right)\eta\right]\,.  \label{ocB}
\end{equation}

When dealing with open Friedmann models, the scale factor and the curvature contribution to the total energy density are conveniently expressed in terms of the conformal time as
\begin{equation}
a= a_0\left[{\sinh(\beta\eta)\over \sinh(\beta\eta_0)}\right]^{1/\beta}  \hspace{10mm} {\rm and} \hspace{10mm} \Omega_K= {1\over(aH)^2}= \tanh^2(\beta\eta)\,, \label{-1FRW}
\end{equation}
respectively. Note that the $\beta$-parameter is decided by the equation of state of the matter and is given by $\beta=(1+3w)/2\neq0$. Here, we will consider the post-inflationary evolution of the universe, which means that $w\geq0$ and $\beta\geq1/2$ always. We are also interested in magnetic fields with super-Hubble correlations. Following Eq.~(\ref{-1FRW}a), if $n$ is the (comoving) eigenvalue of a mode, its (physical) size relative to the Hubble scale is determined by the ratio
\begin{equation}
{\lambda_H\over\lambda_n}= {n\over aH}= n\tanh(\beta\eta)\,,  \label{-1lambdas}
\end{equation}
since $H=a^{\prime}/a^2$.

Let us now confine to marginally open FRW universes. According to Equation~(\ref{-1FRW}b), these spacetimes are characterised by very small values of the conformal time (i.e.,~$\Omega_K\ll1$ implies $\eta\ll1$ and vice versa)~\cite{SS}. It is then straightforward to show that, during the reheating and the dust eras (i.e.,~when $\beta=1/2$), marginally open Friedmann models have $a\propto\eta^2$, $\Omega_K\simeq\eta^2/4$ and $\lambda_H/\lambda_n\simeq n\eta/2$. Throughout the radiation epoch, on the other hand, $\beta=1$ and relations \mbox{(\ref{-1FRW}) and (\ref{-1lambdas})} lead to $a\propto\eta$, $\Omega_K\simeq\eta$ and $\lambda_H/\lambda_n\simeq n\eta$ respectively. All these mean that superhorizon-sized modes in marginally open FRW universes satisfy the constraint $n\eta\ll1$, just like in their spatially flat counterparts. The difference is that now $\eta\ll1$ as well. This ensures that, in marginally open Friedmann models, even small-scale modes with fairly large eigenvalues (i.e.,~with $n\gg1$) can lie outside the Hubble radius (i.e.,~satisfy the condition $n\eta\ll1$).

Superhorizon-sized magnetic fields evolving on spatially open Friedmannian backgrounds obey solution (\ref{hcB}) or (\ref{ocB}), depending on their wavelength (i.e.,~on the range of the associated eigenvalues). On these scales, $n\eta\ll1$ and $\eta\ll1$, when the FRW background is marginally open. Then, both (\ref{hcB}) and~(\ref{ocB}) reduce to the power-law~\footnote{As expected, one can arrive to solution (\ref{-1lscBn}) after evaluating the integration constants of the full solutions (\ref{hcB}) and (\ref{ocB}) and then taking the $(\sqrt{|n^2-2|}\,)\eta\ll1$-limit of the resulting expressions.}
\begin{equation}
\mathcal{B}_{(n)}= a^2B_{(n)}= \mathcal{C}_1+ \mathcal{C}_2\left(\sqrt{|n^2-2|}\right)\eta\,,  \label{-1lscBn}
\end{equation}
as long as
\begin{equation}
\left(\sqrt{|n^2-2|}\,\right)\eta\ll 1\,.  \label{smcon}
\end{equation}
Given that $\eta\ll1$ in marginally open FRW universes, there is an extensive range of wavelengths that satisfy both $n\eta\ll1$ and $(\sqrt{|n^2-2|}\,)\eta\ll1$ at the same time. These include all the modes with $n^2<2$ as well as many having $n^2>2$. For example, a magnetic mode with $n^2=10^2$ lies outside the Hubble radius (i.e.,~has $n\eta\ll1$) and also satisfies condition (\ref{smcon}), as long as $\eta\ll1/10$. In that case, the associated $B$-field evolves according to solution (\ref{-1lscBn}).

We have therefore arrived to an evolution law identical to that of the flat FRW case (compare Equation~(\ref{-1lscBn}) to solution (\ref{lsBn}) in Section~\ref{ssL-SSMA}). Moreover, evaluating the integration constants of (\ref{-1lscBn}) and dropping the mode-index ($n$) for simplicity, we obtain
\begin{equation}
B= \left[B_0-\eta_0\left(2H_0a_0B_0+B_0^{\prime}\right)\right] \left({a_0\over a}\right)^2+ \eta_0\left(2H_0a_0B_0+B_0^{\prime}\right) \left({a_0\over a}\right)^2\left({\eta\over\eta_0}\right)\,,  \label{-1lsB}
\end{equation}
which is identical to solution (\ref{lsB1}). Hence, by simply repeating the process of Section~\ref{ssERRD}, we find that $B\propto a^{-3/2}$ during the reheating and dust eras and $B\propto a^{-1}$ throughout the radiation epoch (see solutions (\ref{rh-dB1}) and (\ref{rB1}) in Section~\ref{ssERRD}). More specifically, recalling that $a\propto\eta^2$ and $H=2/a\eta$ during reheating and dust, we may recast (\ref{-1lsB}) into
\begin{equation}
B= -\left(3B_0-\eta_0B_0^{\prime}\right)\left({a_0\over a}\right)^2+ \left(4B_0+\eta_0B_0^{\prime}\right)\left({a_0\over a}\right)^{3/2}\,. \label{-1dlsB}
\end{equation}

Similarly, when radiation dominates the energy density of the universe, we have $a\propto\eta$ and $H=1/a\eta$. In that case, solution (\ref{-1lsB}) becomes
\begin{equation}
B= -\left(B_0-\eta_0B_0^{\prime}\right)\left({a_0\over a}\right)^2+ \left(2B_0+\eta_0B_0^{\prime}\right)\left({a_0\over a}\right)^{3/2}\,.  \label{-1rlsB}
\end{equation}
All these confirm that magnetic fields on marginally open Friedmann backgrounds can be superadiabatically amplified throughout their post-inflationary evolution. This happens as long as the $B$-fields remain outside the Hubble horizon and the initial conditions allow the second modes on the right-hand side of (\ref{-1dlsB}) and (\ref{-1rlsB}) to dictate the magnetic evolution (see Section~\ref{ssRICs} for further discussion). It is then straightforward to show that the residual comoving strength of such a magnetic field, with current coherence scale around 10~Kpc, is close to $10^{-33}$~G (see Section~\ref{ssFMS} and Equation~(\ref{fB2}) there). This value can increase to $10^{-27}$~G by the time the galaxy is fully formed, while additional growth might possible as well (see Section~\ref{ssCSs} earlier).

Following a series of articles on the magnetic evolution in open FRW universes~\cite{TK}-\cite{BTY}, is was recently claimed that superadiabatic amplification is impossible in marginally open Friedmann models~\cite{SS}. The authors reached this conclusion after arriving at Equation~(\ref{-1lscBn}). Once there, however, they dropped the second mode from the right-hand side of that solution. The authors did so based on the fact that $\eta\ll1$, but without evaluating the integration constants first. As we have shown here, that was not the right decision. This oversight prevented the authors of~\cite{SS} from noticing the importance of their ``redundant'' mode and then from realising that marginally open FRW universes can superadiabatically amplify magnetic fields during their post-inflationary evolution.

Before closing this section, we should briefly comment on the nature of the magnetic modes involved. As we have explained above, solution (\ref{-1lscBn}) applies to magnetic modes with $n^2<2$ and also to those having $n^2>2$, provided they satisfy the conditions $n\eta\ll1$ and $(\sqrt{|n^2-2|}\,)\eta\ll1$ simultaneously. In quantum mechanical terms, modes with $n^2<2$ are termed ``supercurvature'' and they have been claimed to suffer from normalisation problems that make them physically ambiguous. Modes with $n^2>2$, on the other hand, are known as ``subcurvature'' and are physically unambiguous. The range of the subcurvature magnetic modes that experience superadiabatic amplification depends on the value of the conformal time, which in turn is decided by the ``amount'' of spatial curvature. For example, current observations indicate a nearly flat universe with $|\Omega_K|\lesssim10^{-3}$ today. Assuming negative curvature and setting $\Omega_K\simeq10^{-4}$ at present, suggests that $\eta\simeq2\times10^{-2}$ today (see Equation~(\ref{-1FRW}b) and recall that $\beta=1/2$ for dust). Then, at least some of the (subcurvature) magnetic modes with $2<n^2\lesssim50^2$ are currently superadiabatically amplified (i.e.,~they satisfy both $n\eta\ll1$ and $(\sqrt{|n^2-2|}\,)\eta\ll1$ simultaneously). Clearly, as we go back in time the values of $\Omega_K$ and $\eta$ drop significantly, ensuring that many more subcurbature magnetic modes were superadiabatically amplified in the past.

\subsection{Marginally Closed FRW Models}\label{ssMCFRWMs}
Let us now turn our attention to Friedmann models with positive spatial curvature. In terms of conformal time, the scale factor and the curvature contribution to the total energy density of a spatially closed FRW universe are given by
\begin{equation}
a= a_0\left[{\sin(\beta\eta)\over \sin(\beta\eta_0)}\right]^{1/\beta} \hspace{10mm} {\rm and} \hspace{10mm} \Omega_K= -{1\over(aH)^2}= -\tan^2(\beta\eta)\,,  \label{+1FRW}
\end{equation}
respectively. Note that $\beta=(1+3w)/2$ and $w\geq0$, as with the open models discussed in the previous section. Similarly, the relative size of a mode is decided by the ratio $\lambda_H/\lambda_n=n/aH=n\tan(\beta\eta)$. Here, however, the commoving eigenvalue is discrete with $n^2\geq3$. On this background and in the absence of electric currents, magnetic fields obey the differential equation
\begin{equation}
\mathcal{B}^{\prime\prime}_{(n)}+ \left(n^2+2\right)\mathcal{B}_{(n)}= 0\,,  \label{+1cBn}
\end{equation}
which accepts the oscillatory solution
\begin{equation}
\mathcal{B}_{(n)}= \mathcal{C}_1\cos\left[\left(\sqrt{n^2+2}\,\right)\eta\right]+ \mathcal{C}_2\sin\left[\left(\sqrt{n^2+2}\,\right)\eta\right]\,.  \label{+1cB1}
\end{equation}
The above monitors the post-inflationary evolution of large-scale $B$-fields on FRW backgrounds with positive spatial curvature.

Let us now focus upon the marginally closed Friedmann universes. Following (\ref{+1FRW}b), these models are characterised by very small values of the conformal time (i.e.,~$|\Omega_K|\ll1\Leftrightarrow\eta\ll1$), in exact analogy with their marginally open counterparts. In this case we have $a\propto\eta^2$, $\Omega_K\simeq-\eta^2/4$ and $\lambda_n/\lambda_H\simeq n\eta/2$ during the reheating and the dust eras (i.e.,~when $\beta=1/2$). Throughout the radiation epoch, on the other hand, $\beta=1$, $a\propto\eta$, $\Omega_K=-\eta^2$ and $\lambda_n/\lambda_H\simeq n\eta$. Consequently, superhorizon-sized modes on marginally closed Friedmann backgrounds have $n\eta\ll1$ and $\eta\ll1$ simultaneously. Therefore, for magnetic modes that also satisfy the constraint $(\sqrt{n^2+2}\,)\eta\ll1$, solution (\ref{+1cB1}) reduces to the power-law\footnote{Given that $\eta\ll1$ always in marginally closed Friedmann models, it is straightforward to verify that at any given time there is a whole range of eigenvalues that satisfy both $n\eta\ll1$ and $(\sqrt{n^2+2}\,)\eta\ll1$.}
\begin{equation}
\mathcal{B}_{(n)}= a^2B_{(n)}= \mathcal{C}_1+ \mathcal{C}_2\left(\sqrt{n^2+2}\,\right)\eta\,.  \label{+1cB2}
\end{equation}
As with the marginally open Friedmann universes of the previous section, the above is essentially identical to solution (\ref{lsBn}) of the flat FRW models. In fact, after evaluating the integration constants of (\ref{+1cB2}), one recovers solutions (\ref{lsB1}) and (\ref{lsB2}) and all the evolution laws obtained in Section~\ref{ssERRD} earlier. More~specifically, we find that the dominant magnetic mode decays as $B\propto a^{-3/2}$ during both the reheating and the dust eras and like $B\propto a^{-1}$ when radiation dominates the energy density of the universe. Consequently, large-scale $B$-fields in marginally closed Friedmann universes can be superadiabatically amplified throughout their post-inflationary evolution, just like their counterparts in the spatially flat and the marginally open models.

\section{Discussion}\label{sD}
Finding an answer to the question of cosmic magnetism has proved a rather difficult theoretical task. The scenarios of primordial magnetogenesis, in particular, have mainly focused on slowing down the so-called adiabatic magnetic decay and thus increase the residual strength of the seed-field to astrophysically relevant values. So far, almost all of the proposed solutions work outside what we might call ``standard physics''. By breaking away from Maxwellian electromagnetism, for example, it is possible to achieve magnetic magnitudes much larger that the ``conventional'' final strength of $\sim$$10^{-53}$~G. The latter value, however, has been obtained after assuming that primordial magnetic fields decay adiabatically during the whole of their post-inflationary evolution on all scales. Here, we have taken another look into this assumption.

The adiabatic magnetic decay after inflation has been attributed to the high electrical conductivity of the matter during most of reheating and throughout the subsequent eras of radiation and dust. This~has been thought enough to guarantee that the magnetic flux remains conserved at all times and on all scales. Nevertheless, the magnetic flux-freezing cannot be achieved without the electric currents. These currents, however, are formed after inflation by local physical processes and their coherence size can never exceed that of the Hubble horizon. The same is also true for the process of magnetic-flux freezing, which is also causal and therefore it can never affect $B$-fields with superhorizon correlations, without violating causality. In other words, the time required to freeze a superhorizon-sized magnetic field in, is longer than the age of the universe. Therefore, $B$-fields that left the horizon during inflation cannot readjust themselves to their new (post-inflationary) environment and freeze-in, until they have crossed back inside the Hubble radius and have come again into full causal contact. Put another way, applying the ideal-MHD approximation on superhorizon scales violates causality. After all, the ideal-MHD limit is the macroscopic outcome of causal microphysical processes, none of which can affect superhorizon-sized perturbations.

Motivated by the above, we have adopted a current-free treatment, where the magnetic evolution remains unaffected by local physics until the time of horizon re-entry. We found that, as long as they stay outside the Hubble radius, these $B$-fields obey a power-law solution. The latter contains two modes, the second of which decays at a pace slower than the adiabatic after inflation. Depending on the initial conditions, this slowly decaying mode can dominate and thus dictate the post-inflationary magnetic evolution. When this happens superhorizon-sized magnetic fields deplete as $B\propto a^{-3/2}$ throughout the reheating and the dust eras. During the intermediate radiation epoch, on the other hand, the decay rate slows down further to $B\propto a^{-1}$. In general, the ``stiffer'' the equation of state of the matter, the slower the magnetic decay rate.

The initial conditions for the post-inflationary magnetic evolution are set by the field's evolution during the de Sitter phase and by the nature of the transition from inflation to reheating and later to the radiation and the dust eras. Here, following Israel's work on junction conditions, we have employed two qualitatively different but complementary initial-condition scenarios. Alternative approaches are also likely of course. Scenario~A allows for an abrupt change in the background equation of state on the transition hypersurface and for the presence of a thin shell there. Within this scenario, primordial $B$-fields are superadiabatically amplified throughout their post-inflationary evolution as long as they remain outside the Hubble radius. Moreover, for all practical purposes, the amplification occurs irrespectively of the magnetic evolution during the de Sitter phase. This scenario can in principle produce astrophysically relevant $B$-fields, with residual strengths close (or even within) the typical galactic dynamo requirements, without abandoning neither classical electromagnetism nor standard cosmology. Scenario B also allows for a sudden change in the cosmic equation of state, but assumes that there is no thin shell on the matching hypersurface. Here, we found that primordial $B$-fields that decayed adiabatically during the de Sitter regime will continue to do so for the rest of their lifetime. This essentially reproduces the typical scenario of conventional magnetogenesis that leads to astrophysically irrelevant $B$-fields at present. However, scenario B also allows for the superadiabatic amplification after inflation of primordial magnetic fields that did not decay adiabatically in the de Sitter phase. This can have serious implications for the non-conventional mechanisms of cosmic magnetogenesis that amplify their $B$-fields during inflation. More specifically, in connection with the CMB limits on the anisotropy of the universe, scenario~B severely constrains models that achieve relatively strong inflationary amplification for their magnetic fields. On the other hand, when the de Sitter enhancement is mild, scenario~B can help to produce astrophysically promising $B$-fields.

To summarise, causality ensures that there is no a priori flux-freezing on super-Hubble scales, even after inflation. On these wavelengths, primordial $B$-fields are only affected by the background expansion and can be superadiabatically amplified throughout their post-inflationary evolution depending on the initial conditions. Here, we have discussed two simple but complementary initial-condition scenarios. In general, the superadiabatic amplification will (sooner or later) occur, as long as the initial conditions at the start of reheating allow the slowly decaying (i.e.,~the second) modes in solutions (\ref{lsB1}) and (\ref{lsB2}) to survive. Naturally, once back inside the horizon, the electric currents take over and freeze the $B$-fields into the highly conductive cosmic medium. Then onwards, the magnetic flux remains conserved and the adiabatic decay-law is restored.

The aforementioned phase of superadiabatic amplification can increase the residual strength of conventional inflationary produced $B$-fields by many orders of magnitude. For example, a~magnetic mode with current comoving (pre-collapse) scale close to 10~Kpc, which is the minimum required by the galactic dynamo, re-enters the horizon a little before equipartition. This mode has been superadiabatically amplified during reheating and most of the radiation era. As a result, the residual magnetic strength is not the ``standard'' $\sim$$10^{-53}$~G but the much larger $\sim$$10^{-33}$~G. The latter increases further during the protogalactic collapse and can reach strengths within the general dynamo requirements. Additional amplification may also occur during the earlier or the later stages of the field's evolution. Consequently, conventional electromagnetism and standard cosmology can produce magnetic fields of astrophysically relevant magnitudes. The same is also true for the non-conventional scenarios of cosmic magnetogenesis that mildly amplify their $B$-fields during inflation. Those achieving strong amplification during the de Sitter phase, on the other hand, may require revision to avoid potential conflict with the observations.

It is also worth pointing out that our analysis and our results are not confined to the spatially flat Friedmann models, but extend naturally to their marginally open and marginally closed counterparts. Thus, superhorizon-sized magnetic fields of cosmological origin can be superadiabatically amplified in FRW universes with mildly curved (positive or negative) spatial sections as well. To a large extent, this is also intuitively plausible. What is less straightforward and more interesting is that the marginally open models can superadiabatically amplify both supercurvature and subcurvature magnetic modes. The former have been claimed to suffer from normalisation problems that make them physically ambiguous, although the whole issue may merely reflect the absence as yet of a quantum theory of gravity. Nevertheless, the fact that a wide range of the physically unambiguous subcurvature magnetic modes are also superadiabatically amplified, means that the mechanism discussed here works in all the cosmologically relevant Friedmann models

We would like to close with some thoughts on the question of cosmic magnetism and the ongoing efforts to address it. As we have already pointed out, the overwhelming majority of the proposed theoretical solutions operate outside classical Maxwellian theory, or conventional cosmology (or both). In fact, non-conventional magnetogenesis has become a big industry, over the years. This has established in the community the belief that it is not possible to produce cosmological magnetic fields of astrophysical relevance within what we call standard physics. All these are to be expected, to a certain extent at least, since long standing beliefs develop their own inertia as time goes by. Nevertheless, there has been work in the recent literature suggesting that the study of classical electromagnetism on conventional FRW models has not been exhausted yet and the present work takes another step in this direction. The underlying point is that, if a mere appeal to causality can increase the final magnetic strength by 20 or so orders of magnitude, then it might be worth reconsidering the necessity of introducing new physics to address the question of cosmic magnetism.

\end{document}